# Unusual electrical and magnetic properties in layered EuZn$_2$As$_2$

Joanna Blawat[1,2], Madalynn Marshall[3], John Singleton[4], Erxi Feng[5], Huibo Cao[5], Weiwei Xie[3], Rongying Jin[1,*]

[1]Center for Experimental Nanoscale Physics, Department of Physics & Astronomy, University of South Carolina, Columbia, SC 29208, USA

[2]Department of Physics & Astronomy, Louisiana State University, Baton Rouge, LA 70803, USA

[3]Department of Chemistry and Chemical Biology, Rutgers University, Piscataway, NJ 08854, USA

[4]National High Magnetic Field Laboratory, Los Alamos National Laboratory, Los Alamos, New Mexico 87545, USA

[5]Neutron Scattering Division, Oak Ridge National Laboratory, Oak Ridge, TN 37831, USA

**Abstract**

Eu-based compounds often exhibit unusual magnetism, which is critical for nontrivial topological properties seen in materials such as $EuCd_2As_2$. We investigate the structure and physical properties of $EuZn_2As_2$ through measurements of the electrical resistivity, Hall effect, magnetization, and neutron diffraction. Our data show that $EuZn_2As_2$ orders antiferromagnetically with an A-type spin configuration below $T_N$ = 19 K. Surprisingly, there is strong evidence for dominant ferromagnetic fluctuations above $T_N$, as reflected by positive Curie-Weiss temperature and extremely large negative magnetoresistance (MR) between $T_N$ and $T_{fl} \approx 200$ K. Furthermore, the angle dependence of the $MR_{ab}$ indicates field-induced spin reorientation from the *ab*-plane to a direction approximately 45° from the *ab* plane. Compared to $EuCd_2As_2$, the doubled $T_N$ and $T_{fl}$ make $EuZn_2As_2$ a better platform for exploring topological properties in both magnetic fluctuation ($T_N < T < T_{fl}$) and ordered ($T < T_N$) regimes.

1. Introduction

Since the discovery of topological states in semimetals [1–6], the search for new topological materials has been extremely active. Time-reversal symmetry ($\mathcal{T}$) and crystal inversion symmetry ($\mathcal{P}$), together with the Kramers theorem (each energy band is doubly degenerate for fermions), are crucial in understanding the formation of topological states in semimetals[1]. When both $\mathcal{T}$ and $\mathcal{P}$ are preserved, a system may host Dirac fermions at the Dirac points (DP), at which bands are quadruply degenerate. Several Dirac semimetals have been identified, including $Cd_3As_2$ [2], $Na_3Bi$ [3], and $ZrTe_5$ [4]. When $\mathcal{T}$ is broken but $\mathcal{P}$ is preserved, one may expect two Weyl cones with opposite chirality in a Weyl semimetal. Considering the opposite situation, *i.e.*, $\mathcal{T}$ is preserved but $\mathcal{P}$ is broken, the number of Weyl cones is multiplied by four, removing the chirality [1]. Weyl states have been observed in semimetals such as $MoTe_2$ [5,6], $TaAs$ [7], $WTe_2$ [8], and $TaIrTe_4$ [9]. Breaking both $\mathcal{T}$ and $\mathcal{P}$ would generally destroy the Dirac/Weyl properties in these systems. An exception is that the Dirac state can exist if the symmetry of the product $\mathcal{PT}$ is preserved [19]. A case study was carried out in magnetic CuMnAs, which belongs to a type-III magnetic space group with a C-type spin structure [10,11]. Magnetic materials offer a fertile ground for searching for novel topological properties protected by the $\mathcal{PT}$ symmetry.

Recently, materials with the type-IV magnetic space group have also been studied [12]. The type-IV magnetic space group is defined by $G + \mathcal{T}\{e|\tau\}G$, where $G$ is the ordinary nonmagnetic space group, e is the identity operation and $\{e|\tau\}$ represents the translation operation between spin-up and spin-down sublattices in a magnetic space group [12]. In such a magnetic space group, there is a nonsymmorphic time-reversal $\mathcal{T}'$ symmetry, which is related to $\mathcal{T}$ through the translation operation $\tau$ ($\mathcal{T}' = \mathcal{T}\tau$). In the centrosymmetric type-IV magnetic space group, the symmetry of the product $\mathcal{PT}'$ is preserved; therefore Kramers degeneracy is protected, and Dirac Points can exist [12]. EuCd$_2$As$_2$ has been theoretically predicted to be an antiferromagnetic (AFM) Dirac semimetal, in which DPs are protected by the $\mathcal{PT}'$ symmetry [12]. EuCd$_2$As$_2$ forms an A-type AFM state below $T_N$ = 9.5 K in which the threefold symmetry in the *ab*-plane is broken due to the spin configuration. According to theoretical calculations, this leads to a gap between Dirac cones[13,14]. However, strong ferromagnetic (FM) fluctuations above $T_N$ break the $\mathcal{T}$ symmetry, giving rise to a Weyl state at high temperatures ($T > T_N$) [15]. Thus, such a system provides a rare case for studying the transition between the Dirac and Weyl states by tunning temperature. To further study the influence of magnetism on the topology of the electronic band structure, we investigate the physical properties of EuZn$_2$As$_2$, a sister compound of EuCd$_2$As$_2$. The replacement of Cd by Zn is expected to weaken the spin-orbit coupling, thus influencing the gap size between Dirac cones. In addition, compared to EuCd$_2$As$_2$, we find that the AFM ordering temperature for EuZn$_2$As$_2$, $T_N$ = 19 K, is doubled, offering a much wider temperature range for studying potential topological properties in a long-range AFM ordered state. By analyzing magnetization and magneto-transport data, we also find strong evidence for ferromagnetic (FM) fluctuations over a much wider temperature range than that in EuCd$_2$As$_2$. Such information is key toward the understanding of magnetism-driven topological properties.

2. **Experimental methods**

The single crystals of EuZn$_2$As$_2$ were grown via the flux method using Sn. The elemental Eu (99.9% pieces, Alfa Aesar), Zn (99.8% granules, Alfa Aesar), As (99.999% powder, Alfa Aesar) and Sn (99.9% granules, Alfa Aesar) were placed into an alumina crucible with a molar ratio Eu : Zn : As : Sn = 1 : 2 : 2 : 20, and sealed in an evacuated quartz tube. The sample was heated up to 600 °C at a rate of 60 °C/h, and kept at this temperature for 5 hours. This was followed warming to 1000 °C and tempering for 10 hours. The sample was then slowly cooled (-3 °C/h) down to 600 °C with a further centrifuge in order to remove Sn flux. The resulting single crystals have a typical size of 4 mm × 2 mm × 1 mm and are stable in air.

The crystal structure was determined by single crystal X-ray diffraction using a Bruker Apex II single X-ray diffractometer equipped with Mo radiation ($\lambda_{K\alpha}$ = 0.71073 Å), and by powder X-ray diffraction (PXRD) by means of a Rigaku MiniFlex 600 diffractometer with Cu $K_{\alpha 1}$ radiation ($\lambda$ = 1.5406 Å). The crystal structure was solved with the full-matrix least-squares method using the SHELXTL package[16]. The PXRD pattern was analyzed using the FullProf software [17]. The crystal structure is drawn by means of VESTA software[18].

The magnetic properties have been measured in a *Quantum Design* Magnetic Properties Measurement System (MPMS – 7 T). The electrical resistivity, Hall effect, and heat capacity were measured using a *Quantum Design* Physical Property Measurement System (PPMS – 14 T). The standard four-probe technique was used to measure the electrical resistivity and Hall effect, and the relaxation method used for the heat capacity. The angle dependence of the magnetoresistance (MR) was measured at the pulsed-field facility of National High Magnetic Field Laboratory (NHMFL, Los Alamos).

To determine the magnetic structure of EuZn$_2$As$_2$, single-crystal neutron diffraction experiment was performed on DEMAND (HB-3A) at the High Flux Isotope Reactor (HFIR) of Oak Ridge National Laboratory (ORNL)[19]. A wavelength of 1.008 Å from the bent Si-331 monochromator was used to reduce the heavy neutron absorption. Considering the large neutron absorption coefficient of Eu, PLATON software was employed to apply an absorption correction[20,21]. The magnetic and nuclear structures were both determined using Fullprof refinement Suite software[17].

### 3. Results and Discussion

The single crystal X-ray diffraction refinement confirms that our crystals form a trigonal structure with the formula EuZn$_2$As$_2$. The space group is P-*3m1* (No. 164) with the lattice parameters *a* = *b* = 4.2093(1) Å and *c* = 7.175(3) Å. Replacing Cd with smaller Zn, the lattice parameters in EuZn$_2$As$_2$ decrease more significantly along the *a* and *b* axes ($\approx$5.5% reduction) than the *c* axis ($\approx$2.4% reduction). Detailed information including atomic positions and sites occupancies is summarized in Tables I and II. Figure 1(a) illustrates the crystal structure of EuZn$_2$As$_2$, where Zn (green) and As (grey) form a honeycomb network separated by Eu atoms (pink). Figure 1(b) shows the X-ray diffraction pattern of a flat surface of a EuZn$_2$As$_2$ single crystal at room temperature. All peaks can be indexed with the above-mentioned structure from the (001) plane (i.e., the *ab*-plane). A weak peak near 2θ ≈ 30° results from residual Sn on the surface. A picture of a EuZn$_2$As$_2$ single crystal is presented in the inset of Figure 1(b). The PXRD pattern obtained from ground crystals was analyzed using the LeBail method and the result is presented in Figure S1. The red points, black and blue lines represent the experimentally observed intensities, calculated intensities, and the difference between them, respectively. The expected Bragg positions for EuZn$_2$As$_2$ are shown as green vertical marks. The LeBail analysis confirms that EuZn$_2$As$_2$ crystallizes in the above-mentioned trigonal crystal structure.

Figure 2(a) shows the temperature dependence of the zero-field-cooled (ZFC) and field-cooled (FC) magnetic susceptibility measured by applying a magnetic field of 0.1 T parallel to the *c*-axis ($\chi_c$) and to the *ab*-plane ($\chi_{ab}$), respectively. With decreasing temperature, both $\chi_{ab}$ and $\chi_c$ initially increase with little difference between them. Below $T_N$ = 19 K, $\chi_{ab}$ tends to saturate but $\chi_c$ decreases. Such behavior implies that the system forms an A-type AFM order, with FM alignment in the *ab*-plane, but with an AFM configuration along the *c*-axis. The much higher $T_N$ suggests stronger magnetic interactions both within the *ab* plane and along the *c* axis in EuZn$_2$As$_2$ than those in EuCd$_2$As$_2$ [22], consistent with the lattice parameter changes mentioned above.

To understand magnetic interactions in EuZn$_2$As$_2$, we fit $\chi_{ab}$ and $\chi_c$ between 100 K and 300 K using the Curie-Weiss formula $\chi(T) = C/(T - \theta)$, where θ is the Curie-Weiss temperature and *C* is the Curie constant related to the effective magnetic moment $\mu_{eff} = \sqrt{8C}$. As shown in the inset of Figure 2(a), the formula fits the data well (the solid lines are fitting curves). The parameters obtained are $\theta_{ab}$ = 48.3(6) K and $\mu^{ab}_{eff}$ = 7.08(5) μ$_B$, and $\theta_c$ = 25.1(2) K and $\mu^c_{eff}$ = 7.86(3) μ$_B$. The positive values of $\theta_{ab}$ and $\theta_c$ imply dominant ferromagnetic interactions between Eu ions, with the effective magnetic moment close to the theoretical value for Eu$^{2+}$ ($\mu_{eff}$ = 7.94μ$_B$). Taking this at face value, one would expect there to be a difference between the magnetic susceptibility measured under zero-field-cooled (ZFC) and field-cooled (FC) modes; however, this is not seen (Figure 2(a)). Furthermore, the application of higher magnetic field should enhance the FM interactions. Figures 2(b) and 2(c) show the temperature dependences of $\chi_{ab}$ and $\chi_c$ between 2 and 60 K in various fields. Note that increased magnetic field pushes $T_N$ to lower temperatures (Figure 2(d)).

Both observations indicate that EuZn$_2$As$_2$ is *not* a ferromagnet below $T_N$. As shown in Figure 2(e), the field dependence of the magnetization has no hysteresis in either $M_{ab}$ or $M_c$ at 2 K. Instead, both $M_{ab}$(H) and $M_c$(H) vary linearly with field before reaching saturation. The observed saturation moment is close to the theoretical value for Eu$^{2+}$ ($\mu_{sat}$ = gJ = 7$\mu_B$, where J is the total angular momentum and g is the Landé *g* factor)[23].

To confirm the nature of the phase transition at $T_N$, we have measured the temperature and field dependence of the specific heat, $C_p$. Figure 2(f) shows the temperature dependence of $C_p$ at $\mu_0 H$ = 0 and 9 T. Note that there is a lambda-shaped anomaly in $C_p(H=0)$ at $T_N$ = 19 K, indicating a second-order phase transition. The entropy released at the phase transition between 10 and 25 K is $S \approx$ 10.8 J mol$^{-1}$ K$^{-1}$, about 62.5% of the theoretically expected value for an 8-fold degenerate system [$S = R \ln(2J+1) = R \ln 8$ = 17.28 J mol$^{-1}$ K$^{-1}$)]. This reduced entropy is likely related to magnetic fluctuations, which release some amount of entropy above $T_N$. By applying $\mu_0 H$ = 9 T, the specific heat peak is completely suppressed. Notably, $C_p$ ($\mu_0 H$ = 9 T) > $C_p$ ($H$ = 0) above 19 K, confirming partial entropy removal for the case of $H$ = 0 prior to $T_N$.

Given the unusual magnetic behavior seen in the magnetization above and below $T_N$, it is essential to determine the magnetic structure of EuZn$_2$As$_2$ through single-crystal neutron diffraction experiment. Figure 3(a) shows the rocking curve scan at (0 0 ½) peak position at 4 K and 40 K. The peak disappears above $T_N$, indicating a magnetic propagation vector **k** = (0 0 ½) for the ordered state. We find that the magnetic peaks collected at 4 K can be fit by an A-type magnetic structure with the magnetic moments entirely in the *ab*-plane within the data resolution. The resultant magnetic structure at 4 K is presented in Figure 3(b), which is ferromagnetically aligned in the *ab* plane but antiferromagnetic along the *c* axis. This A-type AFM structure is the same as those seen in EuCd$_2$As$_2$ grown with Sn flux [14,22] and in EuMg$_2$Bi$_2$[24] but different from that in EuIn$_2$As$_2$ [25]. At 4 K, the refined magnetic moment is 7.33(7)$\mu_B$/Eu, slightly higher than the saturation value from the magnetization (Figure 2(e)). We choose the reflection intensity (*I*) at (0 0 ½) as the order parameter. The temperature dependence of *I* is shown in Figure 3(c). The solid curve is the fit of *I*(*T*) to $I = A(1 - T/T_N)^{2\beta} + B$, with $T_N \approx$ 19.4 K, $A$ = 3119, $\beta$ = 0.23 and $B$ = 1394. The critical exponent β obtained corresponds to a 2D magnetic system, consistent with the layered structure of EuZn$_2$As$_2$. Figure 3(d) shows the calculated structure factor square ($F^2_{calc}$) versus the observed one ($F^2_{obs}$). The linear behavior indicates excellent structure refinement. Any deviation is likely to be due to the errors resulting from the absorption correction process.

With an A-type magnetic structure (Figure. 3(b)), it is necessary to ask why both $\theta_{ab}$ and $\theta_c$, obtained at temperatures well above $T_N$, are positive. For EuCd$_2$As$_2$, both electron-spin resonance and muon-spin relaxation measurements reveal strong FM fluctuations with long time and length scales, which persist up to approximately 100 K [15]. Although this was not discussed in Ref. [15], we note that the electrical resistivity of EuCd$_2$As$_2$ also begins to show a negative slope below around 100 K. Bearing this in mind, we investigate the temperature and field dependence of both the *ab*-plane ($\rho_{ab}$, $I \parallel ab$) and *c*-axis ($\rho_c$, $I \parallel c$) resistivities. Figure 4(a) shows the temperature dependence of $\rho_{ab}$ and $\rho_c$ between 2 and 300 K for EuZn$_2$As$_2$. Several features are worth mentioning. First, while $\rho_c > \rho_{ab}$ due to the layered structure of EuZn$_2$As$_2$, $\rho_{ab}$ and $\rho_c$ show similar temperature dependence over the entire temperature range. This implies that the scattering mechanism is more or less the same in both the *ab* plane and the *c* direction. Similar behavior has also been reported for EuCd$_2$As$_2$[26]. Second, there is a sharp peak in both $\rho_{ab}$ and $\rho_c$, corresponding to the magnetic transition at $T_N$. The peak in $\rho_c$ is even sharper than that in $\rho_{ab}$. Third, both $\rho_{ab}$ and $\rho_c$ initially vary linearly with temperature at high temperatures, deviating below $T_{fl} \approx 200$ K and eventually acquiring negative slopes (d$\rho_{ab}$/d$T$ < 0 and d$\rho_c$/d$T$ < 0) below ≈150 K. The sharp decrease of $\rho_{ab}$ and $\rho_c$ below $T_N$ indicates that the resistivity in all directions is dominated by spin scattering above $T_N$. The departure from the high-temperature linear behavior marks the spin scattering contribution to the resistivity due to magnetic fluctuations below $T_{fl}$.

To confirm the effect of magnetic fluctuations on $\rho_{ab}$ and $\rho_c$, the magnetic field dependence of $\rho_{ab}$ and $\rho_c$ at constant temperatures is investigated. Figure 4(b) and Figure 4(c) show the field dependence of the transverse ($H \perp I$) magnetoresistivity MR$_{ab}$ ($I \parallel ab$, $H \parallel c$) and MR$_c$ ($I \parallel c$, $H \parallel ab$) at various temperatures between 2 and 300 K, respectively. MR is defined by $MR = \frac{\rho(H)-\rho(H=0)}{\rho(H=0)} \times 100\%$. At 300 K, both MR$_{ab}$ and MR$_c$ are small and positive, typical for a paramagnetic material. Upon cooling, both MR$_{ab}$ and MR$_c$ gradually decrease and become negative near 200 K. With further cooling, their magnitudes continuously increase until $T_N$. The negative MR$_{ab}$ and MR$_c$ indicate the influence of ferromagnetic fluctuation below $T_{fl} \approx 200$ K, consistent with what is seen in EuCd$_2$As$_2$ [15]. Note that $T_{fl}$ for EuZn$_2$As$_2$ is much higher than that in EuCd$_2$As$_2$. Both higher $T_{fl}$ and $T_N$ offer wider temperature ranges for studying magnetism-related properties in EuZn$_2$As$_2$ than in EuCd$_2$As$_2$.

At $T_N$, the spin scattering is almost completely suppressed, so that MR$_{ab}$ and MR$_c$ reach ≈ -90% at $H > H_{sat}$. Below $T_N$, the field dependence of both MR$_{ab}$ and MR$_c$ is non-monotonic, with an initial increase followed by a decrease to negative values [Figures. 4(b) and 4(c)]. The initial positive MR$_{ab}$ and MR$_c$ is attributable to AFM interaction in both the *ab* plane and *c* direction (Figure 3(a))[27]. With increasing field (less than 1 Tesla at 5 K), both MR$_{ab}$ and MR$_c$ start to decrease, eventually becoming negative and saturated above $H_{sat}$. This implies continuous alignment toward the ferromagnetic configuration. When $H$ reaches $H_{sat}$, all moments are aligned ferromagnetically. Figure 4(d) shows the field dependence of MR$_{ab}$ at various angles $\phi$ ($\phi$ = H ^ I - defined in the inset) at $T$ = 0.6 K. The low-field MR$_{ab}$ peak is gradually suppressed as $H$ turns away from the principal axes (*ab*- or *c*-axis). The inset of Figure 4(d) shows MR$_{ab}(\phi)$ at $\mu_0 H$ = 0.5 and 6 T. Note that MR$_{ab}(\phi)$ reaches a minimum around $\phi$ = 45°, implying that there is field-induced spin reorientation pointing along the 45° direction between the *ab*-plane and *c*-axis.

With a strong magnetic fluctuation effect on the resistivity observed, it is interesting to consider the Hall response. Figure 4(e) shows the magnetic field dependence of the Hall resistivity ($\rho_{xy}$) at temperatures between $T_N$ and 300 K. Surprisingly, $\rho_{xy}$ increases linearly with magnetic field at all indicated temperatures, characteristic of the ordinary Hall effect. This means one can extract the Hall coefficient $R_H$ via $\rho_{xy} = \mu_0 R_H H$ for each temperature. Figure 4(f) presents the temperature dependence of $R_H$, which increases with decreasing temperature. The positive $R_H$ suggests that holes are dominant carriers in EuZn$_2$As$_2$. Using the Drude relationship $R_H$ = 1/$ne$, the carrier concentration $n$ can be estimated. As shown in Figure 4(f), $n \approx 4 \times 10^{28}$ cm$^{-3}$ at room temperature, consistent with the semimetallic scenario for EuZn$_2$As$_2$. It increases slightly with increasing temperature, which can be attributed to thermal effect if there is a small band gap as predicted in EuCd$_2$As$_2$. Note that $n$ becomes flat between approximately 100 K and $T_{fl}$, i.e., deviating from its behavior at both high- and low-temperatures. This suggests the influence of FM fluctuations to the band structure. It is our future work to elucidate the connection between magnetic fluctuations and possible topological phase transition in EuZn$_2$As$_2$ as discussed in EuCd$_2$As$_2$ [13-15].

## 4. Summary

We have successfully grown single crystalline EuZn$_2$As$_2$, which forms a trigonal structure. The electrical resistivity, magnetization, and neutron diffraction investigation indicate that EuZn$_2$As$_2$ orders antiferromagnetically below $T_N$ = 19 K with an A-type spin configuration. Similar to EuCd$_2$As$_2$, there are strong FM fluctuations that give rise to profound spin scattering between $T_N$ and $T_{fl}$ ≈ 200 K in EuZn$_2$As$_2$. Our MR$_{ab}$ measurements with variable applied field direction indicate that there is field-induced reorientation of the spins to ≈ 45° between the *ab*-plane and *c*-axis. Compared to EuCd$_2$As$_2$, the doubled $T_N$ and $T_{fl}$ make EuZn$_2$As$_2$ a better platform for exploring topological properties in both magnetic fluctuation ($T_N < T < T_{fl}$) and ordered ($T < T_N$) regimes. It is especially interesting to find out (1) if the reduced spin-orbit coupling in EuZn$_2$As$_2$ will reduce or close the gap between Dirac cones predicted in EuCd$_2$As$_2$, and (2) if the canted spin structure induced by magnetic field impacts the topological states.

## 5. Acknowledgements

Work at University of South Carolina and Louisiana State University is supported by NSF through Grant DMR-1504226. M.M. and W. X. were supported by the U.S. Department of Energy (DOE), Office of Science, Basic Energy Sciences under award DE-SC0022156. E.F. and H.B.C. are supported by the U.S. DOE, Office of Science, Office of Basic Energy Sciences, Early Career Research Program Award KC0402020, under Contract No. DE-AC05-e00OR22725. This research used resources at the High Flux Isotope Reactor and the Spallation Neutron Source, the DOE Office of Science User Facility operated by ORNL. A portion of this work was performed at the National High Magnetic Field Laboratory (NHMFL), which is supported by National Science Foundation Cooperative Agreement No. DMR- 1644779 and the Department of Energy (DOE). JS acknowledges support from the DOE BES program "Science at 100 T", which permitted the design and construction of specialized equipment used in the high-field studies.

## 6. Conflict of interest

Authors declare no conflict of interests.

**Figure 1.** (a) The crystal structure of EuZn$_2$As$_2$. The pink, green and grey balls represent Eu, As and Zn atoms, respectively. (b) The PXRD pattern of EuZn$_2$As$_2$ single crystal. Inset: picture of EuZn$_2$As$_2$ single crystal.

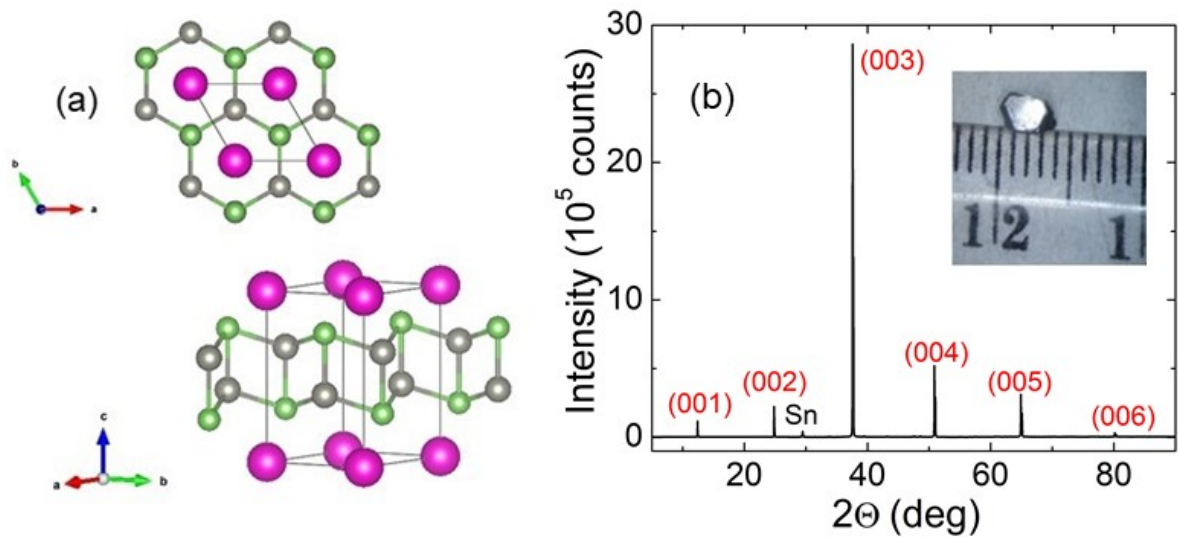

**Figure 2.** (a) The temperature dependence of the magnetic susceptibility measured along ab-plane ($\chi_{ab}$) and c-axis ($\chi_c$). Inset: the inverse magnetic susceptibility measured along ab-plane and c-axis with the Curie-Weiss fit discussed in the text. $\chi_{ab}$ and $\chi_c$ measured at several magnetic fields as a function of temperature (b) and (c), respectively. (d) H – T phase diagram constructed using $\chi_{ab}$ and $\chi_c$ in Figure (b) and (c). (e) Magnetic hysteresis loop measured along ab-plane ($\chi_{ab}$) and c-axis ($\chi_c$). (f) The temperature dependence of specific heat. Inset: the specific heat measured with H = 0 T and 9 T.

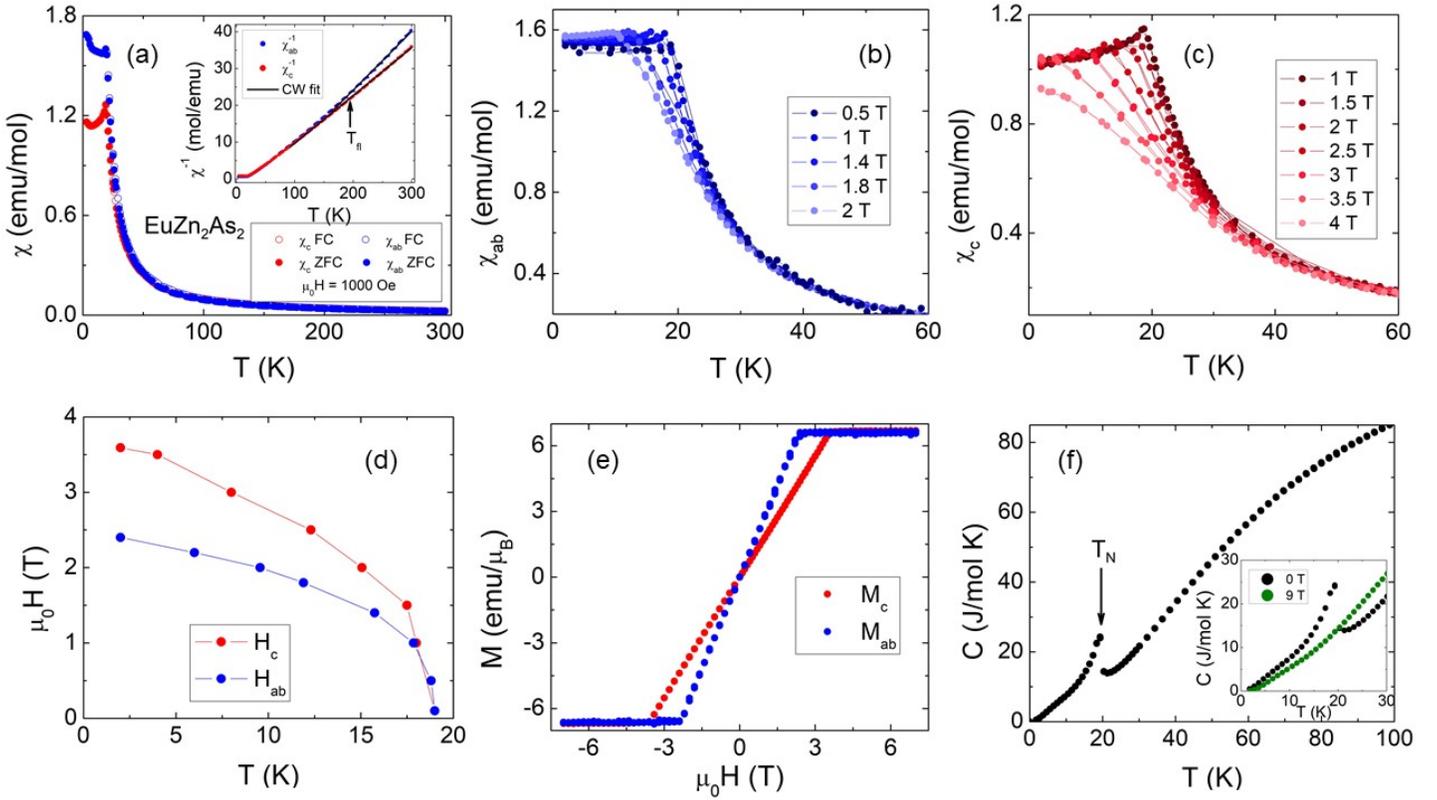

**Figure 3.** (a) The rocking curve scans at (0 0 ½) at 4 K and 40 K, measured by neutrons. (b) magnetic structure of Eu sublattice obtained at T = 4 K. (c) The peak intensity as a function of temperature with the empirical law fit (red solid line) discussed in the text. (d) Calculated structure factor square ($F^2_{calc}$) versus the observed one ($F^2_{obs}$).

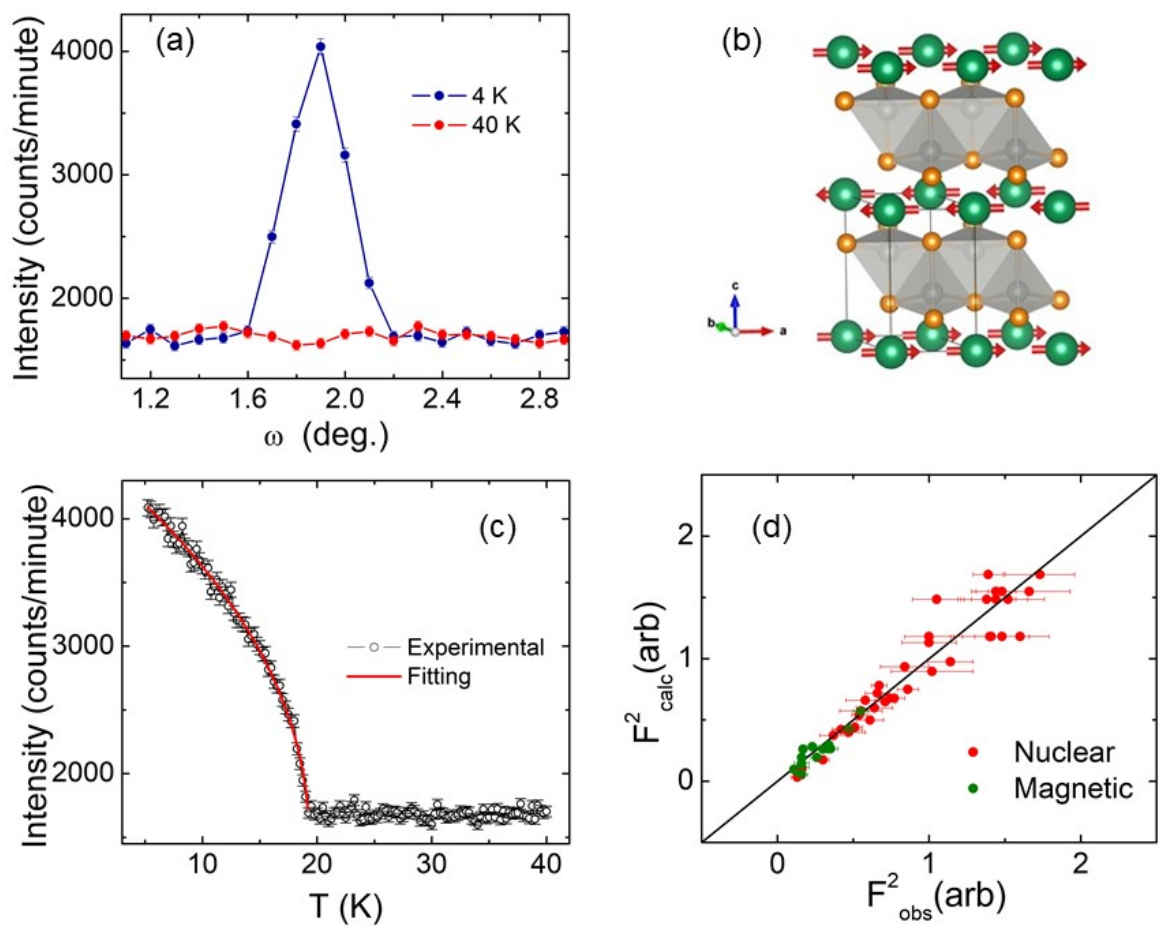

**Figure 4.** (a) Electrical resistivity as a function of temperature measured along ab-plane and c-axis. $MR_{ab}$ (b) and $MR_c$ (c) measured at several temperatures. (d) The magnetic field dependence of magnetoresistivity measured at several different angles. Inset: magetoresistivity versus angle taken at magnetic field $\mu_0H$ = 0.5 T and 6 T. (e) Magnetic field dependence of Hall effect. (f) Hall coefficient and charge density as a function of temperature.

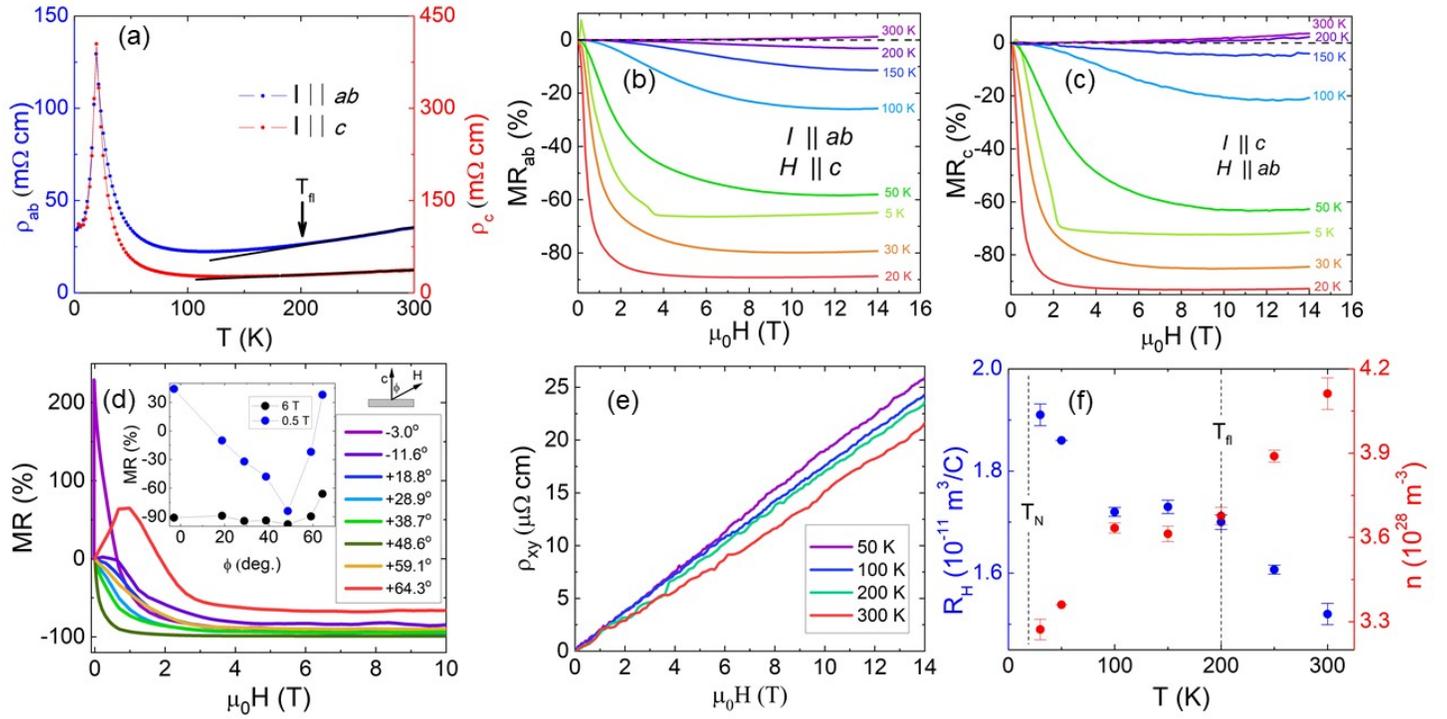

**Table I.** Single crystal crystallographic data and structure refinement for $EuZn_2As_2$.

| Formula | $EuZn_2As_2$ |
|---|---|
| F. W. (g/mol) | 432.54 |
| Space group, Z | $P\text{-}3m1$ (#164), |
| $a$ (Å) | 4.2093(1) |
| $b$ (Å) | 4.2093(1) |
| $c$ (Å) | 7.175(3) |
| V (Å$^3$) | 110.09(6) |
| Absorption correction | Numerical |
| Extinction coefficient | 0.076(4) |
| $\Theta$ range (º) | 2.839 - 33.143 |
| hkl ranges | $-6 \leq h \leq 5$ |
|  | $-6 \leq k \leq 6$ |
|  | $-11 \leq l \leq 10$ |
| No. reflections, $R_{int}$ | 1575, 0.0433 |
| No. independent reflections | 195 |
| No. parameters | 7 |
| $R_1$, $wR_2$ (all $I$) | 0.0258, 0.0378 |
| Goodness of fit | 1.084 |
| Largest diff. peak and hole (e$^-$/Å$^3$) | -1.477, 2.013 |

**Table II.** Atomic coordinates and isotropic displacement parameters of EuZn$_2$As$_2$. $U_{eq}$ is defined as one-third of the trace of the orthogonalized $U_{ij}$ tensor (Å$^2$).

| Atom | Wyckoff | Occupancy | $x$ | $y$ | $z$ | $U_{eq}$ |
|------|---------|-----------|-----|-----|-----|----------|
| Eu | 1b | 1 | 0 | 0 | 0 | 0.0083(2) |
| Zn | 2d | 1 | 1/3 | 2/3 | 0.2667(9) | 0.0074(2) |
| As | 2d | 1 | 1/3 | 2/3 | 0.6296(3) | 0.0106(2) |